\documentclass{JHEP3}
\usepackage{amsmath,amssymb}
\newcommand{\be}{\begin{equation}}
\newcommand{\ee}{\end{equation}}
\newcommand{\ba}{\begin{eqnarray}}
\newcommand{\ea}{\end{eqnarray}}

\def\bi {\begin{itemize}}
\def\ei {\end{itemize}}
\title{On Black Holes and Cosmological Constant in
Noncommutative Gauge Theory of Gravity\footnote{Based on a talk
given at the International Conference on Fundamental and Applied
Research in Physics {\it Farphys 2007}, 25-28 October 2007, Iasi,
Romania.}}

\author{M. Chaichian$^{a}$, M. R. Setare$^{b}$,  A.
Tureanu$^{a}$ and G. Zet$^{c}$\\
$ ^a$High Energy Physics Division, Department of Physical Sciences,
University of Helsinki and Helsinki Institute of Physics, P.O. Box
64, 00014 Helsinki, Finland\\
$ ^b$Department of Science,  Payame Noor University, Bijar, Iran\\
$ ^c$Department of Physics, "Gh. Asachi" Technical University,\\Bd.
D. Mangeron 67, 700050 Iasi, Romania}

\abstract{Deformed Reissner-Nordstr\"om, as well as
Reissner-Nordstr\"om de Sitter, solutions are obtained in a
noncommutative gauge theory of gravitation. The gauge potentials
(tetrad fields) and the components of deformed metric are calculated
to second order in the noncommutativity parameter. The solutions
reduce to the deformed Schwarzschild ones when the electric charge
of the gravitational source and the cosmological constant vanish.
Corrections to the thermodynamical quantities of the corresponding
black holes and to the radii of different horizons have been
determined. All the independent invariants, such as
 the Ricci scalar and the
so-called Kretschmann scalar, have the same singularity structure as
the ones of the usual undeformed case and no smearing of
singularities occurs. The possibility of such a smearing is
discussed. In the noncommutative case we have a local disturbance of
the geometry around the source, although asymptotically at large
distances it becomes flat.}
\begin{document}

\section{Introduction}
\setcounter{equation}{0}

Questioning the nature of space-time at infinitely small scales has
been a fundamental issue for physics. It is generally believed that
the visionary Riemann hinted to a possible breakdown of space-time
as a manifold already in 1854, in his famous inaugural lecture
\cite{Riemann}. The quantum nature of space-time, expressed as
noncommutativity of space-time coordinates, has been lately a
subject of active research, especially in connection with string
theory \cite{6}.

Naturally, various effects of space-time noncommutativity in
cosmology have been studied, principally motivated by the fact that,
since noncommutativity is believed to be significant at the Planck
scale - the same scale where quantum gravity effects become
important - it is most sensible to search for signatures of
noncommutativity in the cosmological observations (for a review, see
\cite{RB} and references therein). One of the most compelling
reasons for the study of noncommutative inflation is the fact that
in an inflationary model the physical wavelengths observed today in
cosmological experiments emerged from the Planckian region in the
early stages of inflation, and thus carry the effects of the Planck
scale physics \cite{ABM}. Among other things, the observed
anisotropies of the cosmic microwave background (CMB) may be caused
by the noncommutativity of space-time \cite{CGS,TMB}. The
noncommutativity has been taken into account either through
space-space uncertainty relations \cite{EGKS} or space-time
uncertainty relations \cite{Brand-Ho}, as well as noncommutative
description of the inflaton (with gravity as background which is not
affected by noncommutativity) \cite{CGS}. On the side of
noncommutative black-hole physics, the studied effect of
noncommutativity was the smearing of the mass-density of a static,
spherically-symmetric, particle-like gravitational source \cite{nic}
(see also \cite{tom}).

These very interesting ideas have been developed lacking a
noncommutative theory of gravity. Although various proposals have
been made (see, for a list of references, \cite{rec}), an ultimate
noncommutative theory of gravity is still elusive. We believe that
the most natural way towards this goal is the gauging of the twisted
Poincar\'e symmetry \cite{10}. Although the formulation of twisted
internal gauge theories has not yet been achieved \cite{13}, the
possibility of gauging the (space-time) twisted Poincar\'e algebra
has not been ruled out and the issue is under investigation.

At the moment, one of the most coherent approaches to noncommutative
gravity is the one proposed by Chamseddine \cite{5}, consisting in
gauging the noncommutative $SO(4,1)$ de Sitter group and using the
Seiberg-Witten map with subsequent contraction to the Poincar\'{e}
(inhomogeneous Lorentz) group $ISO(3,1)$. Although this formulation
is not a final theory of noncommutative gravity, it still can serve
as a concrete model to be studied, whose main features shall
illustrate at least qualitatively the influence of quantum
space-time on gravitational effects. The study of specific examples
as such can cast light upon the reasonable and unreasonable
assumptions proposed so far in the field. Besides, up to now, there
have been no calculations presented in the literature (except
\cite{rec}) to obtain the metric by solving a NC version of
gravitational theory, dues to the technical difficulty of the task.

In a recent paper \cite{rec} a deformed Schwarzschild solution in
noncommutative gauge theory of gravitation was obtained based on
\cite{5}. The gravitational gauge potentials (tetrad fields) were
calculated for the Schwarzschild solution and the corresponding
deformed metric $ \hat{g}_{\mu \nu }\left( {x,\Theta }\right) $ was
defined. According to the result of \cite{rec} corrections appear
only in the second order of the expansion in $\Theta $, i.e. there
are no first order correction terms.

In this paper we attempt to extend the results of \cite{rec} to
include as well the Reissner-Nordstr\"om solution. Having these two
classical solutions known in the noncommutative setup, we can embark
upon a more rigorous study of noncommutative black-hole physics.
Black hole thermodynamical quantities depend on the Hawking
temperature via the usual thermodynamical relations. The Hawking
temperature undergoes corrections from many sources: the quantum
corrections \cite{das}, the self-gravitational corrections
\cite{kk}, and the corrections due to the generalized uncertainty
principle \cite{set6}. In this paper we focus on the corrections due
to the space-space noncommutativity.

The results of this paper have been obtained using a program devised
for GRTensor II application of Maple. For the self-consistence of
the paper we shall present, in the commutative case, results of the
de Sitter gauge theory with spherical symmetry, obtained with a
similar type of program \cite{ze} and recall the derivation of the
metric tensor components in the noncommutative case, as obtained in
\cite{rec}.

\section{de Sitter gauge theory  with spherical symmetry}
\setcounter{equation}{0}

\subsection{Commutative case}

In the following we shall sketch the principal aspects of  a model
of gauge theory for gravitation having the de Sitter group ($dS$) as
local symmetry and gravitational field created by a point-like
source of mass $m$ and carrying also the electric charge $Q$. The
detailed treatment, including the analytical GRTensor II program
used for the calculations, can be found in Ref. \cite{ze}.

The base manifold is a four-dimensional Minkowski space-time
$M_{4}$, in spherical coordinates:
\begin{equation}\label{1}
ds^{2}=-dt^{2}+dr^{2}+r^{2}\left( d\theta ^{2}+\sin ^{2}\theta
d\varphi ^{2}\right) .
\end{equation}
The corresponding metric $g_{\mu \nu }$ has the following non-zero
components:
\begin{equation}\label{2}
g_{00}=-1,\qquad g_{11}=1,\qquad g_{22}=r^{2},\qquad
g_{33}=r^{2}\sin ^{2}\theta .
\end{equation}

The infinitesimal generators of the 10-dimensional de Sitter group
will be denoted by $\Pi _{a}$ and $M_{ab}=-M_{ba}$, $a,b=1,2,3,0$,
where $\Pi _{a}$ generate the de Sitter "translations" and $M_{ab}$
- the Lorentz transformations. In order to give a general
formulation of the gauge theory for the de Sitter group $dS$, we
will denote the generators $\Pi _{a}$ and $M_{ab}$ by $X_{A}$,
$A=1,2,\ldots ,10$. The corresponding 10 gravitational gauge fields
will be the tetrads $ e_{\mu }^{a}(x)$, $a= 0,1,2,3 $, and the spin
connections $\omega _{\mu }^{ab}(x)=-\omega _{\mu }^{ba}(x)$, $[ab]=
[01], [02], [03], [12], [13], [23]$. Then, the corresponding
components of the strength tensor can be written in the standard
form, as the torsion tensor:
\begin{equation}\label{11}
F_{\mu \nu }^{a}=\partial _{\mu }e_{\nu }^{a}-\partial _{\nu }e_{\mu
}^{a}+\left( \omega _{\mu }^{ab}e_{\nu }^{c}-\omega _{\nu
}^{ab}e_{\mu }^{c}\right) \eta _{bc}\,,
\end{equation}
with $\eta_{ab}$ the flat space metric, and the curvature tensor:
\begin{equation}\label{12}
F_{\mu \nu }^{ab}\equiv R_{\mu \nu }^{ab}=\partial _{\mu }\omega
_{\nu }^{ab}-\partial _{\nu }\omega _{\mu }^{ab}+\left( \omega _{\mu
}^{ac}\omega _{\nu }^{db}-\omega _{\nu }^{ac}\omega _{\mu
}^{db}\right) \eta _{cd}+4\lambda ^{2}\left( \delta _{c}^{b}\delta
_{d}^{a}-\delta _{c}^{a}\delta _{d}^{b}\right) e_{\mu }^{c}e_{\nu
}^{d},
\end{equation}%
where $\lambda$ is a real parameter. The integral of action
associated to the gravitational gauge fields $e_{\mu }^{a}(x)$ and
$\omega _{\mu }^{ab}(x)$ will be chosen as \cite{r5}:
\begin{equation}\label{13}
S_{g}=\frac{1}{16\pi G}\int d^{4}x\,e\,F,
\end{equation}
where $e=\det (e_{\mu }^{a})$ and
\begin{equation}\label{14}
F=F_{\mu \nu }^{ab}\,{e}_{a}^{\mu }\,{e}_{b}^{\nu }.
\end{equation}
Here, ${e}_{a}^{\mu }(x)$ denotes the inverse of $e_{\mu }^{a}(x)$
satisfying the usual properties:
\begin{equation}\label{15}
e_{\mu }^{a}{e}_{b}^{\mu }=\delta _{b}^{a},\qquad e_{\mu }^{a}%
{e}_{a}^{\nu }=\delta _{\mu }^{\nu }.
\end{equation}

We assume that the source of the gravitation creates also an
electromagnetic field $A_{\mu }(x)$, with the standard action \cite{r6}:
\begin{equation}\label{16}
S_{em}=-\frac{1}{4Kg^{2}}\int d^{4}x\,e\,A_{\mu }^{a}\,{A}_{a}^{\mu
},
\end{equation}
where $ A_{\mu }^{a}=A_{\mu }^{\nu }e_{\nu }^{a}$, $A_{\mu }^{\nu
}={e}_{a}^{\nu }{e}_{b}^{\rho }\eta ^{ab}A_{\mu \rho }$ and
respectively $ {A}_{a}^{\mu }=A_{\mu }^{\nu }{e}_{a}^{\nu } $, with
$A_{\mu \rho }$ being the electromagnetic field tensor, $A_{\mu \rho
}=\partial _{\mu }A_{\rho }-\partial _{\rho }A_{\mu }$. Here $K$ is
a constant that will be chosen in a convenient form to simplify the
solutions of the field equations and $g$ is the gauge coupling
constant \cite{r6}.

Then, the total integral of action associated to the system composed
of the two fields is given by the sum of the expressions (\ref{13})
and (\ref{16}):
\begin{equation}\label{20}
S=\int d^{4}x\left[ \frac{1}{16\pi G}F-\frac{1}{4Kg^{2}}A_{\mu }^{a}\,%
{A}_{a}^{\mu }\right] e.
\end{equation}
The field equations for the gravitational potentials $e_{\mu
}^{a}(x)$ are obtained by imposing the variational principle $\delta
_{e}S=0$ with
respect to $e_{\mu }^{a}(x)$. They are \cite{r7}:%
\begin{equation}\label{21}
F_{\mu }^{a}-\frac{1}{2}F\,e_{\mu }^{a}=8\pi GT_{\mu }^{a},
\end{equation}%
where $F_{\mu }^{a}$ is defined by:%
\begin{equation}\label{22}
F_{\mu }^{a}=F_{\mu \nu }^{ab}\,{e}_{b}^{\nu },
\end{equation}%
and $T_{\mu }^{a}$ is the energy-momentum tensor of the
electromagnetic
field \cite{r8}:%
\begin{equation}\label{23}
T_{\mu }^{a}=\frac{1}{Kg^{2}}\left( A_{\mu }^{b}\,A_{\nu }^{a}\,{e}%
_{b}^{\nu }-\frac{1}{4}A_{\nu }^{b}\,A_{b}^{\nu }\,e_{\mu
}^{a}\right) .
\end{equation}
The field equations for the other gravitational gauge potentials
$\omega _{\mu }^{ab}(x)$ are equivalent with:
\begin{equation}\label{24}
F_{\mu \nu }^{a}=0.
\end{equation}%

The solutions of the field equations (\ref{21}) and (\ref{24}) were
obtained in  \cite{ze}, under the assumption that the gravitational
field has spherical symmetry and it is created by a point-like
source of mass $m$, which also produces, due to its constant
electric charge $Q$, the electromagnetic field $A_{\mu }(x)$. The
particular form of the spherically symmetric gravitational gauge
field adopted in \cite{ze} is given by the following Ansatz:
\begin{equation}\label{25}
e_{\mu }^{0}=\left( A,0,0,0\right) ,\hspace{0.5cm} e_{\mu }^{1}=\left( 0,\frac{1}{%
A},0,0\right) ,\hspace{0.5cm}e_{\mu }^{2}=\left( 0,0,r,0\right) ,\hspace{0.5cm}%
e_{\mu }^{3}=\left( 0,0,0,r\sin \theta \right) ,
\end{equation}
and
\begin{eqnarray}\label{26}
\omega _{\mu }^{01}=\left( U,0,0,0\right) ,\hspace{0.5cm}\omega
_{\mu }^{02}=\omega _{\mu }^{03}=0,\hspace{0.5cm}\omega _{\mu
}^{12}=\left( 0,0,A,0\right) ,\\ \omega _{\mu }^{13}=\left(
0,0,0,A\sin \theta \right) ,\hspace{0.5cm}\omega _{\mu }^{23}=\left(
0,0,0,\cos \theta \right) ,\nonumber
\end{eqnarray}
where $A$ and $U$ are functions only of the 3D radius $r$. With the
above expressions the components of the tensors $F_{\mu \nu }^{a}$
and $ F_{\mu \nu }^{ab}$ defined by the Eqs. (\ref{11}) and
(\ref{12}) were computed. Here we give only the expressions of $
F_{\mu \nu }^{ab}$ components, which we need to use further, in the
derivation of the expressions of the deformed tetrads:
\begin{eqnarray}\label{29}
  F_{10}^{01} &=&U^{\prime }+4\lambda ^{2},\qquad
F_{20}^{02}=A\left( U+4\lambda ^{2}r\right) ,\qquad
F_{30}^{03}=A\sin \theta \left( U+4\lambda ^{2}r\right) , \cr
F_{21}^{12} &=&\frac{-AA^{\prime }+4\lambda ^{2}r}{A},\qquad
F_{31}^{13}= \frac{\left( -AA^{\prime }+4\lambda ^{2}r\right) \sin
\theta }{A}, \cr
F_{32}^{23} &=&\left( 1-A^{2}+4\lambda ^{2}r^{2}\right) \sin \theta
,
\end{eqnarray}
 where $A^{\prime }$ and $U^{\prime }$ denote the
derivatives with respect to the variable $r$.

Using the field equations, the solution is obtained \cite{ze} as:
\begin{eqnarray}\label{39}
U&=&-AA^{\prime },\cr
 A^{2}&=&1+\frac{\alpha }{r}+\frac{Q^{2}}{r^{2}}+\beta r^{2},
\end{eqnarray}%
where $\alpha $ and $\beta $ are constants of integration. It is
well-known \cite{r8} that the constant $\alpha $ is determined by
the mass $m$ of the point-like source that creates the gravitational
field, by comparison with the Newtonian limit at very large
distances:
\begin{equation}\label{40}
\alpha =-2m.
\end{equation}
The other constant $\beta $ was determined in \cite{ze} (see also
\cite{4}) as $\beta=4\lambda^{2}=-\frac{\Lambda}{3}$, where
$\Lambda$ is the cosmological constant, such that the solution
finally reads:
\begin{equation}\label{com_sol}
A^{2}=1-\frac{2m}{r}+\frac{Q^{2}}{r^{2}}-\frac{\Lambda }{3}r^{2},\qquad U=-%
\frac{m}{r^{2}}+\frac{Q^{2}}{r^{3}}+\frac{\Lambda }{3}r.
\end{equation}

 If we
consider the contraction $\Lambda \rightarrow 0$, then the de Sitter
group becomes the Poincar\'{e} group, and the solution
(\ref{com_sol}) reduces to the Reissner-Nordstr\"{o}m one.

\subsection{Noncommutative case using the Seiberg-Witten map}

The noncommutative corrections to the metric of a space-time with
spherically symmetric gravitational field have been obtained in
\cite{rec}, based on the general outline developed by Chamseddine
\cite{5}.

The noncommutative structure of the space-time is determined by the
commutation relation
\begin{equation}
\left[ {x^{\mu },x^{\nu }}\right] =i\,\Theta ^{\mu \nu },
\label{3.1}
\end{equation}
where $\Theta ^{\mu \nu }=-\,\Theta ^{\nu \,\mu }$ are constant
parameters. It is well known that noncommutative field theory on
such a space-time requires is defined by introducing the star\
product ``*'' between the functions $f$ and $g$ defined over this
space-time:
\begin{equation}
\left( f{\ast g}\right) \left( x\right) =f\left( x\right)
\,e^{\frac{i}{2} \,\Theta ^{\mu\nu }\overleftarrow{\partial_{\mu} }\
\overrightarrow{\partial_{\nu }}}g\left( x\right) . \label{3.2}
\end{equation}

The gauge fields corresponding to the de Sitter gauge symmetry for
the noncommutative case are denoted by  $\hat{e} _{\mu }^{a}\left(
{x,\,\Theta }\right)$ and $\hat{\omega} _{\mu }^{ab}\left(
{x,\,\Theta }\right) $, generically denoted by $\hat{\omega} _{\mu
}^{AB}\left( {x,\,\Theta }\right) $, with the obvious meaning for
the indices $A,B$. The main idea of the Seiberg-Witten map is to
expand the noncommutative gauge fields, transforming according to
the noncommutative gauge algebra, in terms of commutative gauge
fields, transforming under the corresponding commutative gauge
algebra, in such a way that the noncommutative and commutative gauge
transformations are compatible, i.e.
\begin{equation}
\hat{\omega} _{\mu }^{AB}(\omega )+\delta_{\hat\lambda}\hat{\omega}
_{\mu }^{AB}(\omega )=\hat{\omega} _{\mu
}^{AB}(\omega+\delta_\lambda\omega ).
\end{equation}
where $\delta_{\hat\lambda}$ are the infinitesimal variations under
the noncommutative gauge transformations and $\delta_\lambda$ are
the infinitesimal variations under the commutative gauge
transformations.

Using the Seiberg-Witten map \cite{6}, one obtains the following
noncommutative corrections up to the second order \cite{5}:

\begin{eqnarray}
\omega _{\mu \nu \rho }^{AB}\left( x\right)
&=&\frac{1}{4}\,\,\left\{ {\omega _{\nu },\,\partial _{\rho }\omega
_{\mu }+F_{\rho \mu
}}\right\} ^{AB}, \label{3.6}\\
\omega _{\mu \nu \rho \lambda \tau }^{AB}\left( x\right)
&=&\frac{1}{32} \,\,\left( {-\left\{ {\omega _{\lambda },\partial
_{\tau }\left\{ {\omega _{\nu },\partial _{\rho }\omega _{\mu
}+F_{\rho \mu }}\right\} }\right\} +2\left\{ {\omega _{\lambda
},\left\{ {F_{\tau \nu },F_{\mu \rho }}\right\} } \right\} }\right.
\label{3.7}\\
&-&\left\{ {\omega _{\lambda },\left\{ {\omega _{\nu },D_{\rho
}F_{\tau \mu }+\partial _{\rho }F_{\tau \mu }}\right\} }\right\}
-\left\{ {\left\{ { \omega _{\nu },\partial _{\rho }\omega _{\lambda
}+F_{\rho \lambda }} \right\} ,\left( {\partial _{\tau }\omega _{\mu
}+F_{\tau \mu }}\right) } \right\} \cr &+&\left. {2\left[ {\partial
_{\nu }\omega _{\lambda },\partial _{\rho }\left( {\partial _{\tau
}\omega _{\mu }+F_{\tau \mu }}\right) }\right] \,}\right)
^{AB},\nonumber
\end{eqnarray}
where
\begin{equation}
\left\{ {\alpha ,\beta }\right\} ^{AB}=\alpha ^{AC}\,\beta
_{C}^{B}+\beta ^{AC}\,\alpha _{C}^{B},\quad \left[ {\alpha ,\beta
}\right] ^{AB}=\alpha ^{AC}\,\beta _{C}^{B}-\beta ^{AC}\,\alpha
_{C}^{B}  \label{3.8}
\end{equation}
and
\begin{equation}
D_{\mu }F_{\rho \sigma }^{AB}=\partial _{\mu }F_{\rho \sigma
}^{AB}+\left( { \omega _{\mu }^{AC}\,F_{\rho \sigma }^{D\,B}+\omega
_{\mu }^{BC}\,F_{\rho \sigma }^{D\,A}}\right) \,\eta _{CD}.
\label{3.9}
\end{equation}
The noncommutative tetrad fields were obtained in \cite{5} up to the
second order in $\Theta$ in the limit $\Lambda\rightarrow 0$ as:
\begin{equation}
\hat{e}_{\mu }^{a}\left( {x,\Theta }\right) =e_{\mu }^{a}\left(
x\right) -i\,\,\Theta ^{\nu \rho }\,e_{\mu \nu \rho }^{a}\left(
x\right) +\Theta ^{\nu \rho }\,\Theta ^{\lambda \tau }\,\,e_{\mu \nu
\rho \lambda \tau }^{a}\left( x\right) +O\left( \Theta {^{3}}\right)
,  \label{3.10}
\end{equation}
where
\begin{eqnarray}
e_{\mu \nu \rho }^{a}&=&\frac{1}{4}\left[ {\omega _{\nu
}^{a\,c}\partial _{\rho }e_{\mu }^{d}+\left( {\partial _{\rho
}\omega _{\mu }^{a\,c}+F_{\rho \mu }^{a\,c}}\right) \,e_{\nu
}^{d}}\right] \,\eta _{c\,d},  \label{3.11}\\
e_{\mu \nu \rho \lambda \tau }^{a}&=&\frac{1}{32}\left[ {2\left\{
{F_{\tau \nu },F_{\mu \rho }}\right\} ^{a\,b}\,e_{\lambda
}^{c}-\omega _{\lambda }^{a\,b}\left( {D_{\rho }\,F_{\tau \mu
}^{c\,d}+\partial _{\rho }\,F_{\tau \mu }^{c\,d}}\right) \,e_{\nu
}^{m}\,\eta _{d\,m}}\right.\cr
&-&\left\{ {\omega _{\nu },\left( {D_{\rho }F_{\tau \mu }+\partial
_{\rho }F_{\tau \mu }}\right) }\right\} ^{a\,b}\,\,e_{\lambda
}^{c}-\partial _{\tau }\left\{ {\omega _{\nu },\left( {\partial
_{\rho }\,\omega _{\mu }+F_{\rho \mu }}\right) }\right\}
^{a\,b}\,e_{\lambda }^{c}  \label{3.12}\\
&-&\omega _{\lambda }^{a\,b}\,\partial _{\tau }\left( {\omega _{\nu
}^{c\,d}\,\partial _{\rho }e_{\mu }^{m}+\left( {\partial _{\rho
}\,\omega _{\mu }^{c\,d}+F_{\rho \mu }^{c\,d}}\right) \,e_{\nu
}^{m}}\right) \,\eta _{dm}+2\,\partial _{\nu }\omega _{\lambda
}^{a\,b}\partial _{\rho }\partial _{\tau }\,e_{\mu }^{c}\cr
&-&2\,\partial _{\rho }\left( {\partial _{\tau }\,\omega _{\mu
}^{a\,b}+F_{\tau \mu }^{a\,b}}\right) \,\partial _{\nu }\,e_{\lambda
}^{c}-\left\{ {\omega _{\nu },\left( {\partial _{\rho }\omega
_{\lambda }+F_{\rho \lambda }}\right) }\right\} ^{a\,b}\partial
_{\tau }\,e_{\mu }^{c}\cr
&-&\left. {\left( {\partial _{\tau }\,\omega _{\mu }^{a\,b}+F_{\tau
\mu }^{a\,b}}\right) \,\left( {\omega _{\nu }^{c\,d}\partial _{\rho
}e_{\lambda }^{m}+\left( {\partial _{\rho }\,\omega _{\lambda
}^{c\,d}+F_{\rho \lambda }^{c\,d}}\right) \,e_{\nu }^{m}\,\eta
_{d\,m}}\right) }\right] \,\eta _{b\,c}.\nonumber
\end{eqnarray}
Using the hermitian conjugate $\hat{e}_{\mu }^{a\dagger}\left(
{x,\Theta } \right) $\ of the deformed tetrad fields given in
(\ref{3.10}),
\begin{equation}
\hat{e}_{\mu }^{a}{}^{\dagger}\left( {x,\Theta }\right) =e_{\mu
}^{a}\left( x\right) +i\,\,\Theta ^{\nu \rho }\,e_{\mu \nu \rho
}^{a}\left( x\right) +\Theta ^{\nu \rho }\Theta ^{\lambda \tau
}e_{\mu \nu \rho \lambda \tau }^{a}\left( x\right) +O\left( \Theta
{^{3}}\right) . \label{3.13}
\end{equation}
the real deformed metric was introduced in \cite{rec} by the
formula:
\begin{equation}
\hat{g}_{\mu \nu }\left( {x,\Theta }\right) =\frac{1}{2}\,\eta
_{a\,b}\,\left( {\hat{e}_{\mu }^{a}\ast \hat{e}_{\nu
}^{b}{}^{\dagger}+\hat{e} _{\mu }^{b}\ast \hat{e}_{\nu
}^{a}{}^{\dagger}}\right) .  \label{3.14}
\end{equation}

\subsection{Second order corrections to Reissner-Nordstr\"{o}m de
Sitter solution}

Using the Ansatz (\ref{25})-(\ref{26}), we can determine the
deformed Reissner-Nordstr\"{o}m de Sitter metric by the same method
as the Schwarzschild metric was obtained in \cite{rec}. To this end,
we have to obtain first the corresponding components of the tetrad
fields $\hat{e}_{\mu }^{a}\left( {x,\Theta }\right) $ and their
complex conjugated $\hat{e}_{\mu }^{a}{}^{+}\left( {x,\Theta
}\right) $ given by the Eqs. (\ref{3.10}) and (\ref{3.13}). With the
definition (\ref{3.14}) it is possible then to obtain the components
of the deformed metric $\hat{g}_{\mu \nu }\left( {x,\Theta }\right)
$.

Taking only space-space noncommutativity, $\Theta_{0i}=0$ (due to
the known problem with unitarity), we choose the coordinate system
so that the parameters $\Theta ^{\mu \nu }$ are given as:
\begin{equation}
\Theta ^{\mu \nu }=\left(
\begin{array}{cccc}
0 & \Theta & 0 & 0 \\
-\Theta & 0 & 0 & 0 \\
0 & 0 & 0 & 0 \\
0 & 0 & 0 & 0
\end{array}
\right) ,\quad \mu ,\nu =1,2,3,0.  \label{4.1}
\end{equation}

The non-zero components of the tetrad fields $\hat{e}_{\mu
}^{a}\left( {x,\Theta }\right) $ are:
\begin{eqnarray}
\hat{e}_{1}^{1}&=&\frac{1}{A}+\frac{{A}^{\prime \prime
}}{8\,}\,\Theta
^{2}+O( \Theta {^{3}}),  \label{4.2a}\\
\hat{e}_{2}^{1}&=&-\frac{i}{4}\left( {A+2\,r\,{A}^{\prime }}\right)
\,\,\Theta +O( \Theta {^{3}}) ,  \cr
\hat{e}_{2}^{2}&=&r+\frac{1}{32}\,\left( {7A\,{A}^{\prime
}+12\,r\,{A}^{\prime }{}^{2}+12\,r\,A\,{A}^{\prime \prime }}\right)
\,\Theta ^{2}+O( \Theta {^{3}}) ,  \cr
\hat{e}_{3}^{3}&=&r\sin \theta -\frac{i}{4}\left( {\cos \theta
}\right) \Theta +\frac{1}{8}\,\left( {2r\,{A}^{\prime }{}^{2}+r A
{A}^{\prime \prime }+2A{A}^{\prime }-\frac{{A}^{\prime }}{A}}\right)
\left( {\sin \theta }\right) \Theta ^{2}+O(\Theta {^{3}}), \cr
\hat{e}_{0}^{0}&=&A+\frac{1}{8}\,\left( {2\,r\,{A}^{\prime
}{}^{3}+5\,r\,A\,{A} ^{\prime }\,{A}^{\prime \prime
}+r\,A^{2}\,{A}^{\prime \prime \prime }+2\,A\, {A}^{\prime
}{}^{2}+A^{2}\,{A}^{\prime \prime }}\right) \,\,\Theta ^{2}+O(
\Theta {^{3}})\,,  \nonumber
\end{eqnarray}
where ${A}^{\prime },\,{A}^{\prime \prime },\,{A}^{\prime \prime
\prime }$ are first, second and third derivatives of $A(r)$,
respectively, with $A^2$ given in (\ref{com_sol}).

Then, using the definition (\ref{3.14}), we obtain the following
non-zero components of the deformed metric $\hat{g}_{\mu \nu }\left(
{x,\Theta } \right) $\ up to the second order:
\begin{eqnarray}
\hat{g}_{1\,1}\left( {x,\Theta }\right)
&=&\frac{1}{A^{2}}+\frac{1}{4}\,\frac{{ A}^{\prime \prime
}}{A}\,{\Theta }^{2}+O( {\Theta ^{4}}) , \label{4.3}\\
\hat{g}_{22}\left( {x,\Theta }\right) &=&r^{2}+\frac{1}{16}\,\left(
{ A^{2}+11\,r\,A\,{A}^{\prime }+16\,r^{2}\,{A}^{\prime
}{}^{2}+12\,r^{2}A\,{A} ^{\prime \prime }}\right) \,{\Theta }^{2}+O(
{\Theta ^{4}}) , \cr
\hat{g}_{33}\left( {x,\Theta }\right) &=&r^{2}\,\sin ^{2}\,{\theta
}\cr
&+&\frac{1 }{16}\,\left[ {4\,\left( {2\,r\,A\,{A}^{\prime
}-\,r\frac{{A}^{\prime }}{A} +\,r^{2}\,A\,{A}^{\prime \prime
}+2\,r^{2}\,{A}^{\prime }{}^{2}}\right) \,\sin ^{2}\,\theta +\cos
^{2}\theta \,}\right] \,{\Theta }^{2}+O( {\Theta ^{4}})\cr
\hat{g}_{00}\left( {x,\Theta }\right) &=&-A^{2}-\frac{1}{4}\,\left(
{2\,r\,A\,{ A}^{\prime }{}^{3}+r\,A^{3}\,{A}^{\prime \prime \prime
}+A^{3}\,{A}^{\prime \prime }+2\,A^{2}\,{A}^{\prime
}{}^{2}+5\,r\,A^{2}\,{A}^{\prime }\,{A} ^{\prime \prime }}\right)
\,{\Theta }^{2}+O( {\Theta ^{4}}) ,\nonumber
\end{eqnarray}
For $\Theta \rightarrow 0$ we obtain the commutative
Reissner-Nordstr\"{o}m de Sitter solution with $A^{2}=1-\frac{2m
}{r}+\frac{Q^{2}}{r^{2}}-\frac{\Lambda}{3}r^2$.

We should mention that the expressions for the noncommutative
corrections to the deformed tetrad fields and noncommutative metric
elements are the same as the ones obtained in \cite{rec}, although
here we have also the electromagnetic field involved. The reason is
that the first order noncommutative corrections to the
electromagnetic field in the Seiberg-Witten map approach, i.e.
$A_\mu^{(1)}$ is a pure gauge and thus can be gauged away
\cite{Jiang}. As a result, in this order noncommutative corrections
involving the electromagnetic field do not appear.

Now, if we insert $A$ into (\ref{4.3}), then we obtain the deformed
Reissner-Nordstr\"{o}m-de Sitter metric with corrections up to the
second order in $\Theta$. Its non-zero components are:
\begin{eqnarray}
\hat{g}_{11}&=&\left(1-\frac{2m
}{r}+\frac{Q^{2}}{r^{2}}\right)^{-1}+\frac{(-2mr^3+3m^2r^2+3Q^2r^2-6mQ^2r+2Q^4)}{16r^2(r^2-2mr+Q^2)}\Theta^{2}\,,\label{4.5}\\
\hat{g}_{22}&=&r^2+\frac{r^4-17mr^3+34m^2r^2+27Q^2r^2-75mQ^2r+30Q^4}{16r^2(r^2-2mr+Q^2)}\Theta^{2}\,,\cr
\hat{g}_{33}&=&r^2\sin^2\,\theta +\frac{\cos^{2}\theta( r^4+2
mr^3-7Q^2r^2-4 m^2 r^2+16m
  Q^2 r-8  Q^4)}{16r^2(r^2-2mr+Q^2)}\Theta ^{2}\cr
  &+&\frac{(-4mr^3+4m^2r^2+8Q^2r^2-16mQ^2r+8Q^4)
 }{16r^2(r^2-2mr+Q^2)}\Theta ^{2}\,,\cr
\hat{g}_{00}&=&-\left(1-\frac{2m
}{r}+\frac{Q^{2}}{r^{2}}\right)+\frac{4mr^3-9Q^2r^2-11m^2r^2+30mQ^2r-14Q^4}{4r^6}\Theta
^{2}\,.\nonumber
\end{eqnarray}

\subsection{Noncommutative scalar curvature and cosmological constant}

It is well known that, in the commutative case, the scalar curvature
of the vacuum solutions (like the Schwarzschild and
Reissner-Nordstr\"om, when $\Lambda=0$) vanishes, i.e. the
corresponding space-time is Ricci-flat. It is interesting to study
whether this property holds also in the noncommutative case.
Moreover, this study is motivated also by the fact that in the
commutative case the addition of a cosmological term leads to
nonvanishing scalar curvatures even in the space devoid of any
gravitational source. Should the scalar curvature not vanish for the
deformed vacuum solutions, the noncommutative behaviour, in some
sense or another, may naturally imitate the commutative solution
with cosmological term.

The noncommutative Riemann tensor is expanded in powers of $\Theta$
as \cite{5}:
\begin{equation}\label{rim}
\hat{F}^{ab}_{\mu\nu}=F^{ab}_{\mu\nu}+i\Theta^{\rho\tau}F^{ab}_{\mu\nu\rho\tau}+\Theta^{\rho\tau}\Theta^{\kappa\sigma}
F^{ab}_{\mu\nu\rho\tau\kappa\sigma}+O(\Theta^{3})\,,
\end{equation}
where
\begin{equation}\label{rim1}
F^{ab}_{\mu\nu\rho\tau}=\partial_{\mu}\omega^{ab}_{\nu\rho\tau}+(\omega^{ac}_{\mu}\omega^{db}_{\nu\rho\tau}+
\omega^{ac}_{\mu\rho\tau}+\omega^{db}_{\nu}-\frac{1}{2}\partial_{\rho}\omega_{\mu}^{ac}\partial_{\tau}\omega_{\nu}^{db})
\eta_{cd}-(\mu\leftrightarrow \nu)
\end{equation}
and
\begin{equation}\label{rim2}
F^{ab}_{\mu\nu\rho\tau\kappa\sigma}=\partial_{\mu}\omega^{ab}_{\nu\rho\tau\kappa\sigma}+
(\omega^{ac}_{\mu}\omega^{db}_{\nu\rho\tau\kappa\sigma}+
\omega^{ac}_{\mu\rho\tau\kappa\sigma}+\omega^{db}_{\nu}-\omega^{ac}_{\mu\rho\tau}\omega^{db}_{\nu\kappa\sigma}-
\frac{1}{4}\partial_{\rho}\partial_{\kappa}\omega^{ac}_{\mu}\partial_{\tau}\partial_{\sigma}\omega_{\nu}^{db})\eta_{cd}-(\mu\leftrightarrow
\nu)\,,
\end{equation}
where $\omega^{ab}_{\mu\nu\rho}$ and
$\omega^{ab}_{\mu\nu\rho\lambda\tau}$ are given by Eqs. (\ref{3.6})
and (\ref{3.7}), respectively, with $A=a$ and $B=b$. After we
calculate $\hat{F}^{ab}_{\mu\nu}$ and have also $\hat{e}^{a}_{\mu}$
we can obtain the noncommutative scalar curvature:
\begin{equation}\label{rim3}
\hat{F}=\hat{e}^{\mu}_{ a}\ast \hat{F}^{ab}_{\mu\nu}\ast
\hat{e}^{\nu}_{b}
\end{equation}
where $\hat{e}^{\mu}_{a}$  is the inverse of $\hat{e}^{a}_{\mu}$
with respect to the star product, i.e. $\hat{e}^{\mu}_{
a}\ast\hat{e}^{b}_{\mu}=\delta^b_a$. The general expression of the
scalar curvature, expanded in powers of $\Theta$, is:
\begin{eqnarray}\label{rim4}
\hat{F}=F&+&\Theta^{\rho\tau}\Theta^{\kappa\sigma}(e^{\mu}_{a}F^{ab}_{\mu\nu\rho\tau\kappa\sigma}e^{\nu}_{b}+
e^{\mu}_{a\rho\tau\kappa\sigma}F^{ab}_{\mu\nu}e^{\nu}_{b}+e^{\mu}_{a}F^{ab}_{\mu\nu}e^{\nu}_{b\rho\tau\kappa\sigma}
-e^{\mu}_{a\rho\tau}F^{ab}_{\mu\nu}e^{\nu}_{b\kappa\sigma}\cr
&-&e^{\mu}_{a\rho\tau}F^{ab}_{\mu\nu\kappa\sigma}e^{\nu}_{b}
-e^{\mu}_{a}F^{ab}_{\mu\nu\rho\tau}e^{\nu}_{b\kappa\sigma})+O(\Theta^{4})\,.
\end{eqnarray}
Here, we have to calculate also $e^{\mu}_{a\rho\tau}$ and
$e^{\mu}_{a\rho\tau\kappa\sigma}$, using:
\begin{equation}\label{rim5}
\hat{e}^{\mu}_{
a}=e^{\mu}_{a}-i\Theta^{\nu\rho}e^{\mu}_{a\nu\rho}+\Theta^{\nu\rho}\Theta^{\kappa\sigma}e^{\mu}_{a\nu\rho\kappa\sigma}
+O(\Theta^{3})\,.
\end{equation}

The noncommutative scalar curvature for the Reissner-Nordstr\"{o}m
de Sitter solution is then obtained in the form:
\begin{eqnarray}\label{rim7}
\hat{F}&=&4\Lambda+\frac{96m r^5-552m^2 r^4-72Q^2 r^4+896m^3
r^3+1174m Q^2 r^3-2740m^2 Q^2 r^2}{32r^8(r^2-2m r+Q^2)}\Theta^{2}\cr
&+&\frac{-550Q^4 r^+22362m Q^4 r-614 Q^6}{32r^8(r^2-2m
r+Q^2)}\Theta^{2}\cr &+&\frac{(-16m^2 r^4+28Q^2 r^4-24m Q^2 r^3+12
Q^4 r^2)}{32r^8(r^2-2m r+Q^2)}(\cot^{2}\theta)\ \Theta^{2}
\end{eqnarray}
For the charge $Q=0$ one obtains the scalar curvature for the
deformed Schwarzschild de Sitter solution, which is also non-zero.

The very interesting feature of these scalar curvatures is that for
finite values of the radius $r$ they are non-zero for the pure
Schwarzschild and Reissner-Nordstr\"om vacuum solutions (when
$\Lambda=0$), although asymptotically they do vanish. In a away,
this situation locally mimics the presence of a nonvanishing
cosmological term in the Einstein equations, where it is well known
that the space-time curvature does not vanish even in the absence of
any matter.

\section{Noncommutativity corrections to the thermodynamical quantities of black holes}
\setcounter{equation}{0}

In this section we derive the corrections to the  thermodynamical
quantities due to the space-space noncommutativity.

For the Reissner-Nordstrom-de Sitter metric, there are four roots of
$A^{2}(r)=1-\frac{2m }{r}+\frac{Q^{2}}{r^{2}}-\frac{\Lambda}{3}r^2$,
denoted $r_i$, with $i=1,...,4$. In the Lorentzian section, $0\leq
r<\infty$, the first root is negative and has no physical
significance. The second root $r_2$ is the inner (Cauchy) black-hole
horizon, $r = r_3=r_+$ is the outer (Killing) horizon, and $r =
r_4=r_c$ is the cosmological (acceleration) horizon. The solution of
\begin{equation}\label{cos0}
A^{2}(r)=1-\frac{2m }{r}+\frac{Q^{2}}{r^{2}}-\frac{\Lambda}{3}r^2=0
\end{equation}
is found as a series expansion in the cosmological constant:
\begin{equation}\label{cos}
r=r_0+a\Lambda +b \Lambda^{2}+...\,,
\end{equation}
where $r_0$ is the Reissner-Nordstr\"om horizon radius because for
$\Lambda=0$ we obtain the Reissner-Nordstr\"om solution. We know
that:
\begin{equation}\label{cos1}
r_0=m\pm \sqrt{m^2-Q^2}\,.
\end{equation}
Therefore, we obtain the following cosmological and black-hole
horizon radius solutions with cosmological constant, respectively:
\begin{equation}\label{cos2}
r_c=m+ \sqrt{m^2-Q^2}+\frac{(m+ \sqrt{m^2-Q^2})^{5}}{6(m(m+
\sqrt{m^2-Q^2})-Q^2)}\Lambda\,,
\end{equation}
\begin{equation}\label{cos3}
r_+=m- \sqrt{m^2-Q^2}+\frac{(m+\sqrt{m^2-Q^2})^{5}}{6(m(m+
\sqrt{m^2-Q^2})-Q^2)}\Lambda\,.
\end{equation}

On the other hand, for the Schwarzschild-de Sitter metric, with
$A^{2}(r)=1-\frac{2m }{r}-\frac{\Lambda}{3}r^2$, the number of
positive roots (and thus the number of event horizons) depends on
the ratio of $m$ and $\frac{8\pi}{\Lambda}$. There are no event
horizons if $m>\frac{1}{3\sqrt{\Lambda}}$. There is only one event
horizon, $r_1=\frac{1}{\sqrt{\Lambda}}$, if
$m=\frac{1}{3\sqrt{\Lambda}}$. In the case
$m<\frac{1}{3\sqrt{\Lambda}}$,  there are two distinct event
horizons:
\begin{equation}\label{h2}
r_2=\frac{2}{\sqrt{\Lambda}}\cos
\left(\frac{\pi}{3}+\frac{1}{3}\arctan \sqrt{\frac{1}{9m^2
\Lambda}-1}\right)\,,
\end{equation}
\begin{equation}\label{h3}
r_3=\frac{2}{\sqrt{\Lambda}}\cos
\left(\frac{\pi}{3}-\frac{1}{3}\arctan \sqrt{\frac{1}{9m^2
\Lambda}-1}\right)\,.
\end{equation}
It can be shown that $r_2< r_1< r_3$.

In the noncommutative case, we consider the corrected event horizon
radius up to the second order as
\begin{equation}\label{41}
\hat{r}_{1,2}=A_{1,2}+B_{1,2}\Theta+C_{1,2}\Theta^{2}\,.
\end{equation}
Substituting the above $\hat{r}_{1,2}$ into the equation
$\hat{g}_{00}=0$, we obtain the corrected cosmological and black
hole (Killing) event horizon radii respectively as solutions of this
equation:
\begin{eqnarray}\label{42}
\hat{r}_{1}&=&m+ \sqrt{m^2-Q^2}+\frac{(m+
\sqrt{m^2-Q^2})^{5}}{6(m(m+ \sqrt{m^2-Q^2})-Q^2)}\Lambda\cr
&+&\frac{(6m^4+\sqrt{m^{2}-Q^2}(6m^3-8mQ^2)-11Q^2m^2+5Q^4)}
{8(8m^5+\sqrt{m^{2}-Q^2}(8m^4-8m^2Q^2+Q^4)-12m^3Q^2+4mQ^4)}\Theta^{2}
\end{eqnarray}
and
\begin{eqnarray}\label{421}
\hat{r}_{2}&=&m- \sqrt{m^2-Q^2}+\frac{(m+\sqrt{m^2-Q^2})^{5}}{6(m(m+
\sqrt{m^2-Q^2})-Q^2)}\Lambda\cr
&+&\frac{(6m^4-\sqrt{m^{2}-Q^2}(6m^3-8mQ^2)-11Q^2m^2+5Q^4)}
{8(8m^5-\sqrt{m^{2}-Q^2}(8m^4-8m^2Q^2+Q^4)-12m^3Q^2+4mQ^4)}\Theta^{2}
\end{eqnarray}
The distance between the corrected event horizon radii is given by
following relation in an example case, when $m=2Q$
\begin{equation}\label{422}
\hat{d}=\hat{r}_{1}-\hat{r}_{2}=d-\Delta
d=2\sqrt{3}Q+\frac{51\sqrt{3}}{4Q}\Theta^{2}
\end{equation}
Therefore in the noncommtative space-time the distance between
horizons is more than in the commutative case. Then we obtain from
(\ref{422}):
\begin{equation}\label{423}
\frac{\Delta d}{d}=\frac{51 \Theta^{2}}{8Q^2}
\end{equation}
The ratio of this change due to the noncommutativity correction to
the distance has a value which is much too small.

  The modified
Hawking-Bekenstein temperature and the horizon area of the
Reissner-Nordstr\"{o}m de Sitter black hole in noncommutative
space-time to the second order in $\Theta$ are as follows,
respectively:
\begin{eqnarray}\label{43}
\hat{T}_+&=&\frac{1}{4\pi}\frac{d\hat{g}_{00}(\hat{r}_{1})}{
dr}=\frac{m^2-m\sqrt{m^2-Q^2}-Q^2}{2\pi(m-\sqrt{m^2-Q^2})}\\
&+&\frac{Q^2(-4m^2Q^2\sqrt{m^2-Q^2}-48m^5
+68m^3Q^2-21mQ^4+\sqrt{m^2-Q^2}Q^4)}{12\pi(m-\sqrt{m^2-Q^2})^{4}(-m^2-m\sqrt{m^2-Q^2}+Q^2)}\Lambda\cr
&+&(\frac{(448m^9-1648Q^2m^7+2112Q^4m^5-1091Q^6m^3+179Q^8m)}{16\pi(m+\sqrt{m^{2}-Q^2})^{7}
[8m^5-12Q^2m^3+4Q^4m+\sqrt{m^{2}-Q^2}(8m^4-8Q^2m^2+Q^4)] }\cr
&+&\frac{\sqrt{m^{2}-Q^2}(-2240Q^2m^6+2597Q^4m^4-1053Q^6m^2+612m^8+84Q^8)}{16\pi(m+\sqrt{m^{2}-Q^2})^{7}
[8m^5-12Q^2m^3+4Q^4m+\sqrt{m^{2}-Q^2}(8m^4-8Q^2m^2+Q^4)]}\cr
&+&\frac{(m^2-Q^2)^{3/2}(264Q^2m^4-473Q^4m^2-152m^6+51Q^6)}{16\pi(m+\sqrt{m^{2}-Q^2})^{7}
[8m^5-12Q^2m^3+4Q^4m+\sqrt{m^{2}-Q^2}(8m^4-8Q^2m^2+Q^4)]} \cr
&+&\frac{(m^2-Q^2)^{5/2}(16Q^2m^2-12m^4)}{16\pi(m+\sqrt{m^{2}-Q^2})^{7}
[8m^5-12Q^2m^3+4Q^4m+\sqrt{m^{2}-Q^2}(8m^4-8Q^2m^2+Q^4)]})
\Theta^{2}\,,\nonumber
\end{eqnarray}
\begin{eqnarray}\label{44}
\hat{A}_+=4\pi\hat{r}_{1}^{2}&=&4\pi((m-\sqrt{m^{2}-Q^2})^2+\frac{(m-\sqrt{m^{2}-Q^2})(m+\sqrt{m^2-Q^2})^{5}}{3(m(m+
\sqrt{m^2-Q^2})-Q^2)}\Lambda)\\
&+&\frac{\pi(m+\sqrt{m^{2}-Q^{2}})[6m^4-11Q^2m^2+5Q^4+\sqrt{m^{2}-Q^{2}}(6m^3-8Q^2m)]}{8m^5-12Q^2m^3+4Q^4m+
\sqrt{m^{2}-Q^{2}}(8m^4-8Q^2m^2+Q^4)}\Theta^{2}\,.\nonumber
\end{eqnarray}
The corresponding quantities for the cosmological horizon are as
follows, respectively:
\begin{eqnarray}\label{431}
\hat{T}_c&=&\frac{-1}{4\pi}\frac{d\hat{g}_{00}(\hat{r}_{2})}{
dr}=\frac{-m^2-m\sqrt{m^2-Q^2}+Q^2}{2\pi(m+\sqrt{m^2-Q^2})}\\
&+&\frac{(-4m^2-4m)\sqrt{m^2-Q^2}+5Q^2)(m+\sqrt{m^2-Q^2})}{12\pi(-m^2-m\sqrt{m^2-Q^2}+Q^2)}\Lambda\cr
&+&(\frac{(448m^9-1648Q^2m^7+2112Q^4m^5-1091Q^6m^3+179Q^8m)}{16\pi(m+\sqrt{m^{2}-Q^2})^{7}
[8m^5-12Q^2m^3+4Q^4m+\sqrt{m^{2}-Q^2}(8m^4-8Q^2m^2+Q^4)] }\cr
&+&\frac{\sqrt{m^{2}-Q^2}(-2240Q^2m^6+2597Q^4m^4-1053Q^6m^2+612m^8+84Q^8)}{16\pi(m+\sqrt{m^{2}-Q^2})^{7}
[8m^5-12Q^2m^3+4Q^4m+\sqrt{m^{2}-Q^2}(8m^4-8Q^2m^2+Q^4)]}\cr
&+&\frac{(m^2-Q^2)^{3/2}(264Q^2m^4-473Q^4m^2-152m^6+51Q^6)}{16\pi(m+\sqrt{m^{2}-Q^2})^{7}
[8m^5-12Q^2m^3+4Q^4m+\sqrt{m^{2}-Q^2}(8m^4-8Q^2m^2+Q^4)]} \cr
&+&\frac{(m^2-Q^2)^{5/2}(16Q^2m^2-12m^4)}{16\pi(m+\sqrt{m^{2}-Q^2})^{7}
[8m^5-12Q^2m^3+4Q^4m+\sqrt{m^{2}-Q^2}(8m^4-8Q^2m^2+Q^4)]} )
\Theta^{2}\,,\nonumber
\end{eqnarray}

\begin{eqnarray}\label{441}
\hat{A}_c=4\pi\hat{r}_{2}^{2}&=&4\pi ((m+\sqrt{m^2-Q^2})^2+\frac{(m+
\sqrt{m^2-Q^2})^{6}}{3(m(m+ \sqrt{m^2-Q^2})-Q^2)}\Lambda)\\
&+&\frac{\pi(m+\sqrt{m^{2}-Q^{2}})[6m^4-11Q^2m^2+5Q^4+\sqrt{m^{2}-Q^{2}}(6m^3-8Q^2m)]}{8m^5-12Q^2m^3+4Q^4m+
\sqrt{m^{2}-Q^{2}}(8m^4-8Q^2m^2+Q^4)}\Theta^{2}\,.\nonumber
\end{eqnarray}
According to the Bekenstein-Hawking formula the thermodynamic
entropy of a black hole is proportional to the area of the event
horizon $S=A/4$, where $A$ is the area of the horizon. The corrected
entropy due to noncommutativity for the black-hole horizon and the
cosmological horizon are:

\begin{eqnarray}\label{45}
\hat{S}_+=\frac{\hat{A}_+}{4}&=&\pi^{2}((m-\sqrt{m^{2}-Q^2})^2+\frac{(m-\sqrt{m^{2}-Q^2})(m+\sqrt{m^2-Q^2})^{5}}{3(m(m+
\sqrt{m^2-Q^2})-Q^2)}\Lambda)\\
&+&\frac{\pi\Theta^{2}(m+\sqrt{m^{2}-Q^{2}})[6m^4-11Q^2m^2+5Q^4+\sqrt{m^{2}-Q^{2}}(6m^3-8Q^2m)]}{8m^5-12Q^2m^3+4Q^4m+
\sqrt{m^{2}-Q^{2}}(8m^4-8Q^2m^2+Q^4)}\cr
&+&\frac{\pi\Theta^{2}(m+\sqrt{m^{2}-Q^{2}})[6m^4-11Q^2m^2+5Q^4+\sqrt{m^{2}-Q^{2}}(6m^3-8Q^2m)]}{4[8m^5-12Q^2m^3+4Q^4m+
\sqrt{m^{2}-Q^{2}}(8m^4-8Q^2m^2+Q^4)]}\nonumber
\end{eqnarray}
\begin{eqnarray}\label{451}
\hat{S}_c=\frac{\hat{A}_c}{4}&=&\pi^{2}((m+\sqrt{m^2-Q^2})^2+\frac{(m+
\sqrt{m^2-Q^2})^{6}}{3(m(m+ \sqrt{m^2-Q^2})-Q^2)}\Lambda)\\
&+&\frac{\pi\Theta^{2}(m+\sqrt{m^{2}-Q^{2}})[6m^4-11Q^2m^2+5Q^4+\sqrt{m^{2}-Q^{2}}(6m^3-8Q^2m)]}{8m^5-12Q^2m^3+4Q^4m+
\sqrt{m^{2}-Q^{2}}(8m^4-8Q^2m^2+Q^4)}\cr
&+&\frac{\pi\Theta^{2}(m+\sqrt{m^{2}-Q^{2}})[6m^4-11Q^2m^2+5Q^4+\sqrt{m^{2}-Q^{2}}(6m^3-8Q^2m)]}{4[8m^5-12Q^2m^3+4Q^4m+
\sqrt{m^{2}-Q^{2}}(8m^4-8Q^2m^2+Q^4)]}\nonumber
\end{eqnarray}
If we consider the $Q=0$ case, we obtain the corresponding
quantities for the Schwarzschild de Sitter black holes.

\section{Conclusions and discussions}
\setcounter{equation}{0}

Following Ref. \cite{ze}, in the present paper we constructed a
gauge theory for gravitation using the de Sitter group as the local
symmetry. The gravitational field has been described by gauge
potentials. The solutions of the gauge field equations were studied
considering a spherically symmetric case. Assuming that the source
of the gravitational field is a point-like mass electrically
charged, we obtained the Reissner-Nordstr\"{o}m solution. Then, a
deformation of the gravitational field has been performed along the
lines of Ref. \cite{5} by gauging the noncommutative de Sitter
$SO(4,1)$ group and using the Seiberg-Witten map. The corresponding
space-time is also of Minkowski type, but endowed now with spherical
noncommutative coordinates. We determined the deformed gauge fields
up to the second order in the noncommutativity parameters $\Theta
^{\mu \nu }$. The deformed gravitational gauge potentials (tetrad
fields) \^{e}$_{\mu }^{a}\left( { x,\Theta }\right) $ have been
obtained by contracting the noncommutative gauge group $SO(4,1)$ to
the Poincar\'{e} (inhomogeneous Lorentz) group $ISO(3,1)$. Then, we
have calculated these potentials for the case of the
Reissner-Nordstr\"{o}m  solution and defined the corresponding
deformed metric $ \hat{g}_{\mu \nu }\left( {x,\Theta }\right) $. By
finding the Reissner-Nordstr\"{o}m solution, as well as the
Schwarzschild solution in \cite{rec}, for a noncommutative theory of
gravity we came closer to plausible black-hole physics on
noncommutative space-time. The event horizon of the black hole
undergoes corrections from the noncommutativity of space as in Eq.
(\ref{42}). Since the noncommutativity parameter is small in
comparison with the length scales of the system, one can consider
the noncommutative effect as perturbations of the commutative
counterpart. Then we have obtained the corrections to the
temperature and entropy given in Eqs. (\ref{43}) and (\ref {44}).

The noncommutativity of space-time drastically changes the topology
of the space-time in the vicinity of the source in the presence of
gravitational fields, in the sense that the curvature is not zero,
locally, while asymptotically is does vanish. This situation is, in
a limited sense, similar to the effect of a nonvanishing
cosmological term in usual Einstein's equations. It could not be a
priori ruled out that in a fully consistent treatment of a
noncommutative theory of gravity, without expansion in $\Theta$, the
effects of the cosmological constant could be less locally imitated
by the noncommutativity. In any case, one can say that the NC
corrections are of the same form as those arising from the quantum
gravity effects \cite{khriplovich}.

The use of the Seiberg-Witten map for constructing the
noncommutative gauge theory of gravity leads inevitably to some loss
of information, at least concerning the "big picture", i.e. the
global features of the space-time. The reason is that the
compatibility of the commutative and noncommutative gauge
transformation is required at algebraic level, for infinitesimal
transformations. As a result, the noncommutative fields and,
consequently, the observables, will always be expressed as power
series in $\Theta$, starting from the 0th order, which is inevitably
the corresponding field or observable of the commutative theory.
Thus, the Seiberg-Witten map approach is useful for calculating
corrections, but some phenomena which may be peculiar to the entire
noncommutative setting will be concealed. The phenomenon of UV/IR
mixing \cite{UVIR} is a show-case for this. If we did perturbation
in $\Theta$, the nonplanar diagram (which is finite when using the
whole star-product) would no more be finite and the planar diagram
would remain UV-divergent.

This features of the Seiberg-Witten map may hide interesting aspects
when it comes to singularities. In this paper we have obtained the
same singularity structure for the Schwarzschild and
Reissner-Nordstr\"om metrics: if the 0th order in $\Theta$ is
singular, then higher order corrections could never cancel this
singularity.\footnote{This phenomenon of disappearance of
singularities in the noncommutative case reveals itself in the
noncommutative instantons and solitons as nonperturbative solutions
\cite{nekrasov}.}. This is valid for the deformed Ricci scalar
curvature, as well as the Kretschmann invariant, $\hat
F^{\mu\nu\rho\sigma}\hat F_{\mu\nu\rho\sigma}$, where $\hat
F^{\mu\nu}_{\rho\sigma}$ is the deformed Riemann tensor. This is in
sharp contradiction with the conclusions of Ref. \cite{nic}, where a
nonsingular de Sitter geometry was found in the origin. In Ref.
\cite{nic} the noncommutativity is taken into account by one of its
major effects, the infinite nonlocality which it produces - the
source of gravitational field is not point-like, but it has a
Gaussian extension, while the noncommutativity effects of the
gravitational field have not been taken into account. A clear-cut
conclusion can be provided only by a full treatment of the
noncommutative theory of gravity.

\vskip 0.5cm {\bf{Acknowledgements}}

We are indebted to Archil Kobakhidze, Sergey Odintsov and Shin
Sasaki for useful discussions.

\end{document}